\documentclass[reprint, amsmath, amssymb, aps, prb]{revtex4-2}

\usepackage[utf8]{inputenc}
\usepackage{amsmath}
\usepackage{mathtools}
\usepackage{braket}
\usepackage{graphicx}
\usepackage{subfig}
\usepackage{appendix}
\usepackage{float}
\usepackage{dsfont}
\usepackage{xcolor}
\usepackage{comment}
\usepackage{tikz}
\usepackage{ragged2e}

\DeclareUnicodeCharacter{0308}{\"{}}

\newcounter{remark}

\def \Technion{The Physics Department and the Solid State Institute,
Technion--Israel Institute of Technology,\newline 32000 Haifa, Israel}
\def \Canada{National Research Council of Canada, Ottawa, Ontario, Canada K1A 0R6}

\begin{document}

\title{Temporal CW polarization-tomography of photon pairs from the biexciton radiative cascade: theory and experiment}

\author{Noam Tur$^{1}$}
\author{Ismail Nassar$^{1}$}
\author{Ido Schwartz$^{1}$}
\author{Joseph Avron$^{1}$}
\author{Dan Dalacu$^{2}$}
\author{Philip J. Poole$^{2}$}
\author{David Gershoni$^{1}$}

\affiliation{$^1$\Technion}
\affiliation{$^2$\Canada}

\date{\today}

\begin{abstract}
We study, experimentally and theoretically, temporal correlations between the polarization of photon pairs emitted during the biexciton-exciton radiative cascade from a single semiconductor quantum dot, optically excited by a continuous-wave light source.  
The system is modeled by a Lindbladian coupled to two Markovian baths: One bath represents the continuous light source, and a second represents the emitted radiation. 
Very good agreement is obtained between the theoretical model that we constructed and a set of 36 different time resolved, polarization correlation measurements between cascading photon pairs. 

\end{abstract}

\maketitle

\section{Introduction}\label{s:intro}

The temporal correlations between the polarization states of two photons emitted during the biexciton-exciton radiative cascade in a single semiconductor quantum dot (QD) have been the subject of many studies during the last three decades \cite{regelman, benson, akopian, muller2014demand}. These studies were motivated by the quest to find technological sources for entangled photons, which QDs, also known as 'artificial atoms' \cite{kastner1993artificial, ashoori1996electrons} are expected to form \cite{benson,freedman1972experimental, aspect1981experimental}.

Unlike excited atoms, however, where the fundamental optical excitation is typically Kramers' degenerate, the fundamental optical excitation of  QDs - the electron-hole pair (or exciton) is typically non-degenerate due to the anisotropic exchange interaction between the electron and the hole \cite{gammon, kulakovskii, gupalov1998fine}. 
The anisotropic exchange interaction is due to asymmetry between the electron and hole envelope wavefunctions. 
This asymmetry is due to deviation of the long range exchange interaction from a $C_3v$ symmetry expected from the (111) crystallographic growth direction of the sample (see section II below) \cite{ivchenko2012superlattices}. The deviation is most likely due to  the QD's spatial composition fluctuations, strain fluctuations and/or misalignment between the symmetry axis of the QD and the (111) growth direction \cite{zielinski2015}.

The degeneracy removal of the optically active two level exciton system, reflects itself in temporal dependence of the correlations between the polarization states of the sequentially emitted cascading photons \cite{winik}. As a result, the expected measured entanglement between the polarization states of the first (biexciton-) photon and the second (exciton-) photon is rather small, and its detection typically requires either spectral \cite{akopian} or temporal \cite{winik, stevenson2008evolution}, post-selection.   

In these experiments, two types of optical or electrical \cite{ulrich2003triggered, stevenson2008evolution, huber2014polarization} excitations are used. The first, experimentally simpler to perform, but more difficult to analyze, is to use a continuous wave (CW)  source \cite{akopian, stevenson2008evolution}.  The second, experimentally more challenging, but rather straight forward to analyze \cite{ulrich2003triggered, huber2014polarization, winik, schmidgall2015all, pennacchietti2024oscillating} is to use a periodic short pulse source for the excitation.

To the best of our knowledge, a comprehensive model for the first case has not been developed yet. Therefore, experimental data analysis has so far relied on sometimes partially justified assumptions. 

In this work, we discuss and develop a theoretical model for the CW excitation case. Though the mathematical formulation is somewhat abstract, the developed model is rather easy to encode and to compute. 
The model and code that we developed is thoroughly discussed in this paper and then compared with experimentally measured time resolved polarization sensitive two-photon correlations. 

The paper is organized as follows:
In section II we describe the experimental system, in section III we discuss and develop the theoretical model. When the model development requires more detailed mathematical tools we send the reader to the Appendices. In section IV we present model simulations and accurately fit the developed model to the time resolved polarization-tomography measurements. Section V is a short summary of the paper.

\section{Experimental setup}

The studied sample  contained single InAsP quantum dots embedded in InP nanowires. The sample fabrication method is described in detail in previous publications \cite{dalacu2009selective, dalacu2012ultraclean, bulgarini2014nanowire, cogan2018depolarization, laferriere2021systematic}. 
In brief, the growth was on a $\text{SiO}_{2}$ patterned $(111)$B InP substrate consisting of circular holes opened up in the oxide mask. Gold was deposited in these holes using a self-aligned lift-off process, which allows the nanowires to be positioned at known locations on the substrate. The growth had  two steps: (i) growth of the nanowires' cores containing the QDs, nominally 500 nm from the nanowires' bases, and (ii) cladding of the core to realize nanowire diameters of around 200 nm for efficient light extraction. The QDs' diameters are determined by the size of their cores. 
The particular QD reported on here has a diameter of about 20 nm.

\begin{figure}[ht]
\begin{center}
    \centering
    \subfloat[]{\includegraphics[width=1\columnwidth]{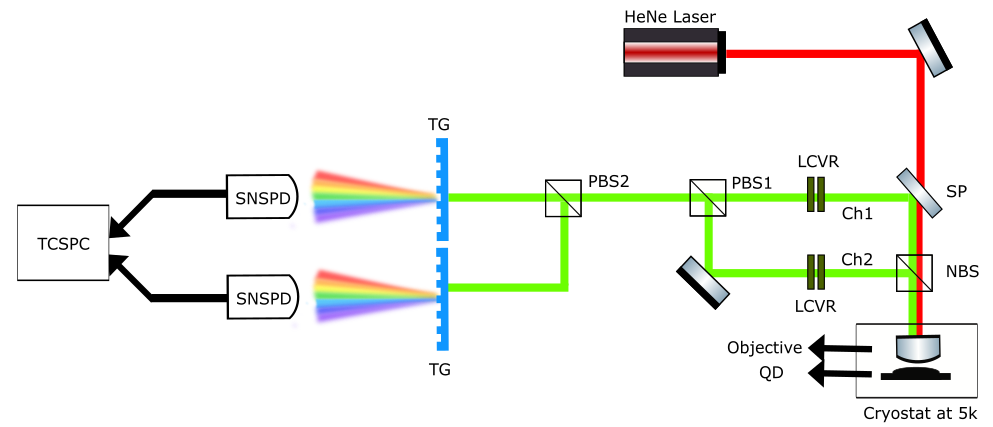}}
    \hfill
    \subfloat[]{\includegraphics[width=1\columnwidth]{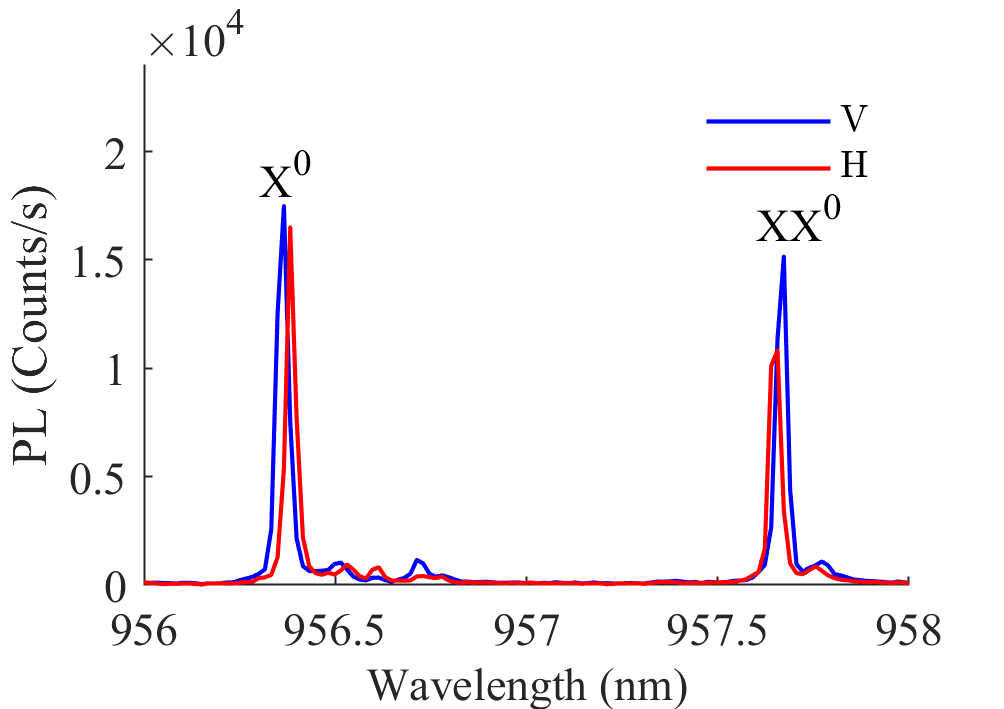}}
    \caption{\justifying (a) Schematic description of the experimental setup. The sample, held at $\approx 5$K in a cryostat. A Short-pass (SP) filter transmits the exciting HeNe laser beam (red solid line) and reflects the collected PL (green solid line). A 0.85 NA objective, focuses the exciting beam and collects the PL.  A non-polarizing beam splitter (NPBS) divides the PL  into two channels. In each channel a pair of liquid crystal variable retarders (LCVRs) is used for polarization projection of the emitted PL onto one polarizing beam splitter (PBS). The PL is then spectrally filtered by pairs of transmission gratings (TGs) and detected by superconducting nanowire single-photon detectors (SNSPDs). 
    The detected events are recorded by a time-correlated single-photon counting (TCSPC) module.
    (b) Rectilinear polarization-sensitive photoluminescence spectra of the biexciton-exciton radiative cascade under CW excitation intensity in which the two spectral lines are nearly equal.
    }\label{f:experiment}
\end{center}
\end{figure}

The experimental setup is shown in Fig.~\ref{f:experiment}(a). The sample was maintained at $\approx 5$K inside a cryostat. CW excitation was provided by a 632.8 nm HeNe laser, focused on a single nanowire using an objective lens with a numerical aperture of 0.85, which also collected the emitted photoluminescence (PL). The emitted PL was split into two channels using a non-polarizing beam splitter (NPBS). A short-pass filter in one channel separated the transmitted excitation light from the reflected PL. Both channels then passed through pairs of liquid crystal variable retarders (LCVRs), projecting the light’s polarization onto a polarizing beam splitter (PBS). This combined horizontal polarization from one channel with vertical polarization from the other. The PL was spectrally filtered using a transmission grating, achieving spectral resolution of 0.02 nm, and then detected by superconducting nanowire single-photon detectors. The detectors provided temporal resolution of about 40 ps, with system overall light harvesting efficiency of about 1-2\%. Finally, the detected events were recorded using a time-tagging single-photon counter.

The rectilinear polarization-sensitive PL spectra from the QD under CW excitation intensity in which the exciton ($X^0$ at 956.4 nm) and biexciton (${X\!X}^0$ at 957.7 nm) spectral lines are nearly equal are shown in Fig.~\ref{f:experiment}(b). 

\section{Theoretical Model}
\subsection{The system}
In the experiment the biexciton and the exciton photons are collected, spectrally filtered, and their polarization is projected before their (random) detection time is registered. We denote by $P_1$ the polarization projection of the biexciton photon and by $P_2$ the polarization projection of the exciton photon. We performed 36 different measurements for $6 \times 6$ pairs of cascading photon polarization combinations in which $P_1,P_2\in\{H,V,D,\bar{D},R,L\}$, where $H(V)$ denotes horizontal- (vertical-) rectilinear polarization, $D (\bar{D})$ diagonal-(anti-diagonal-) linear polarization and $R(L)$ right-(left-) hand circular polarization.

In each measurement, the (random) detection times of the biexciton and exciton photons are recorded. Then the (random) time-difference $\tau$ between temporally close biexciton and exciton photon detection events is stored. We note that $\tau$ can be negative when the exciton photon is detected prior to the biexciton photon. 
The data is then presented as 36 histograms where in each histogram the number of measured $P_1-P_2$ polarized biexciton-exciton correlation events in a given temporal bin $(\tau- \delta\tau/2, \tau +\delta\tau/2)$ are displayed. 
These normalized histograms form the measured polarization sensitive intensity correlation function \cite{scully}:  
\begin{equation}
\label{e:intensity-correlation}
    g^{(2)}_{XX_{P_1}-X_{P_2}}\left(\tau\right) = \frac{\braket{N_{XX_{P_1}}\left(t \right) \cdot N_{X_{P_2}}\left(t+\tau \right) }_t}{\braket{N_{XX_{P_1}}\left(t \right)}_t \cdot \braket{N_{X_{P_2}}\left(t \right)}_t}
\end{equation}
where $N_{XX_{P_1}}\left(\tau \right)$ ($N_{X_{P_2}}\left(\tau \right)$) is the number of detected  $P_1$ ($P_2$) polarized biexciton (exciton) photons during the time bin $\tau\pm\delta\tau$, and the averaging is over the time $t$. Recall that $g^{(2)}_{XX_{P_1}-X_{P_2}}\left(\tau=\pm\infty\right) =1$ as distant detection events are independent.

\begin{figure}[h] 
\begin{center}
  \begin{tikzpicture}[scale=10/4]
    \draw [ultra thick,shift={(0,0)}](-1/2,.8) -- (1/2,.8) node [right] {$\ket{X\!X}$};
    \draw [ultra thick,shift={(0,0)}](-1/2,-.6) -- (1/2,-.6)  node [right] {$\ket{0}$};
    \draw [ultra thick,shift={(0,0.05)}](-1/4,-.05) -- (-4/4,-.050)  node [left] {$\ket{X_H}$};
    \draw [ultra thick,shift={(0,.05)}](1/4,0) -- (4/4,0)  node [right] {$\ket{X_V}$};
    \draw [red,thick,->] (-3/4,0)--(0,-.55) node [left] at (-4/8,-.35) {$H_2$};
     \draw [blue,thick,->] (3/4,0)--(0.1,-.55) node [right] at (1/2,-.35) {$V_2$};
    \draw [blue,thick,->, shift={(3/4,9/12)}] (-3/4,0)--(0,-.7) node [right] at (-2/8,-.35) {$V_1$};
    \draw [red,thick,->, shift={(3/4,9/12)}] (-3/4,0)--(-6/4,-.7) node [left] at (-4/3,-.35) {$H_1$};
    \draw [<->] (0,-.051) -- (0,.051) node [ above] {$\Delta$};
    \end{tikzpicture}
\end{center} 
 \caption{\justifying Schematic description of the biexciton-exciton radiative cascade. $H(V)$ denotes horizontal (vertical)-rectilinear polarization and the subscript (1 or 2) the temporal order of emission. $\ket{X\!X}$ denotes the biexciton state and $\ket{X_H}/\ket{X_V}$ are the 2 bright exciton eigenstates. $\Delta$ (greatly exaggerated in the figure) is the fine structure energy splitting (FSS) between the exciton eigenstates. $\ket 0$ denotes empty QD state. 
}\label{f:cascade}
 \label{fig:my_label}
\end{figure}
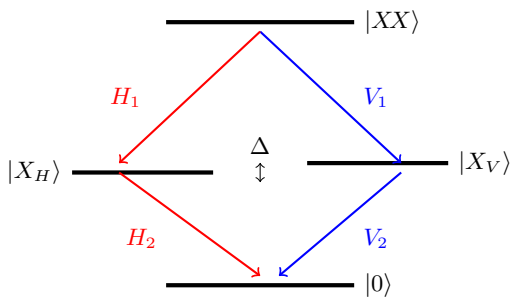

The biexciton-exciton cascade is shown schematically in Fig.~\ref{f:cascade}. As shown in Fig.~\ref{f:cascade}, the system contains a biexciton $\ket{XX}$ state that during a radiative recombination of one of its two e-h pairs emits a single photon. The photon detection heralds a state in which a single exciton occupies the QD. 
The biexciton-exciton optical selection rules are such that the exciton polarization state is determined or heralded by the polarization of the emitted biexciton-photon. For example: detection of an H polarized biexciton photon heralds the exciton in the $\ket{X_H}$ eigenstate and detection of a $V$ polarized biexciton photon heralds the exciton in the $\ket{X_V}$ eigenstate.
Detection of a biexciton photon in any other polarization base, heralds the exciton in a coherent superposition of both of its eigenstates. 
The exciton (second e-h pair) can then recombine radiatively by emitting a photon whose polarization matches the polarization state of the exciton at the recombination time, leaving the QD empty ($\ket{0}$) and thereby completing the radiative cascade. 

The emitted two photons during the radiative cascade are entangled in their energy and polarization degrees of freedom \cite{akopian}.

In modeling the measured intensity correlation functions, several approaches have been considered so far. The first model as in Ref. \cite{regelman}, for example, describes the population dynamics of the various exciton states using a set of first order rate equations. This rate equation model is straightforward and very efficient computationally. It was successfully used for accurately fitting the measured data in Ref. \cite{regelman}, since due to the lack of temporal resolution, the coherence between the two exciton eigenstates, could be ignored. 
However, for understanding the measured high resolution temporal dependence of the polarization tomography of the intensity correlation function, this coherence must be accurately considered.

Another, more advanced model, as in \cite{winik}, for example, employs a Hamiltonian formalism, which is robust for modeling closed quantum systems.  This model has been used successfully in Ref. \cite{winik} for describing the measured polarization sensitive intensity correlation functions in the case of periodic pulsed excitation, which can be described as a closed quantum system \cite{scully}.

\begin{figure}[h]
  \centering
  \begin{tikzpicture}
      \draw[thick,shift={(-3,0)}] (-1.1,-1.1) rectangle (1.1,1.1);
 \draw[thick,shift={(3,0)}] (-1.1,-1.1) rectangle (1.1,1.1);
 \draw [red] (0,0) circle [radius=1];
 \node [shift={(-3,0)}] at (0,0) {radiation}; 
  \node [shift={(3,0)}] at (0,0) {electron-hole}; 
   \node [red] at (0,0) {QD}; 
   \draw [->,thick, red] (-1.2,0)--(-1.8,0);
   \draw [<->,thick, black] (1.2,0)--(1.8,0);
  \end{tikzpicture}
  \caption{\justifying  QD as an open system interacting with a bath of electron-hole pairs in the hosting crystal and a photon bath that absorbs the infrared  photons emitted by the QD. } \label{f:baths}
\end{figure}
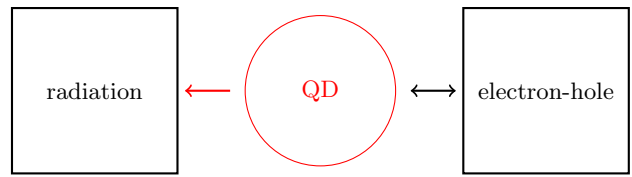

In contrast, under CW excitation the biexciton-exciton cascade is better described as an open system due to its steady state interaction with the environment \cite{scully} which we describe as a bath of electron and hole pairs and another bath which absorbs the emitted photons, as schematically shown in Fig. \ref{f:baths}. 

To consider these baths in the model, we replaced the rate equations in a set of Lindblad equations \cite{lindblad,gorini,davies,bp,abolfath2013dynamical}. The model that we constructed this way effectively accounts for both the coherent evolution of the exciton in the QD and its interactions with the environment, making it particularly well-suited for the steady state conditions that the system reaches under the CW excitation.

\subsection{Lindbladian Model}
The solution to the Lindblad equation \cite{lindblad} 
\begin{eqnarray}\label{e:lindblad-S}
    \frac{d\boldsymbol{\rho} }{d t}&=&{\cal L}(\boldsymbol{\rho}) \nonumber \\
    {\cal L}(\boldsymbol{\rho})&=& {\cal H}(\boldsymbol{\rho}) +\sum_{ij} {\cal D}_{ij}(\boldsymbol{\rho})
\end{eqnarray}
provides a description of the temporal evolution of the system. In this equation the system is represented by a density matrix $\boldsymbol{\rho}$ composed of all the system's states,
and the Lindbladian operator $\mathcal{L}$, is composed of  the Hamiltonian ${\cal H}$ which describes the unitary evolution of the closed system and the operators ${\cal D}_{ij}$ which describe 
the transition from state $j$ to $i$ caused by the interaction with the environment.

Explicitly
\begin{eqnarray}\label{e:superHsuperD}
    {\cal H}(\boldsymbol{\rho})&=& -\frac{i}{\hbar}[\mathbf{H},\boldsymbol{\rho}] \nonumber \\
    {\cal D}_{ij}(\boldsymbol{\rho})&=& \frac{1}{2}[\boldsymbol{\Gamma}_{ij} \boldsymbol{\rho}, \boldsymbol{\Gamma}_{ij} ^\dagger] +\frac{1}{2} [\boldsymbol{\Gamma}_{ij} ,\boldsymbol{\rho} \boldsymbol{\Gamma}_{ij} ^\dagger]
\end{eqnarray}
where $\mathbf{H}$ is the Hamiltonian of the system and $\boldsymbol{\Gamma}_{ij}$ are Lindblad jump operators  \cite{lindblad}.

We proceed by introducing the projection operator $\boldsymbol{\Pi}_{XX}$,$\boldsymbol{\Pi}_{0}$, and $\boldsymbol{\Pi}_{X\left(\theta,\phi \right)}$, which projects the density matrix on the biexciton state, on the empty QD state and on the coherent superposition of the exciton eigenstates 
\begin{equation} \label{e:exciton_superposition}
    \ket{X\left(\theta,\phi \right)} = \cos{\left(\frac{\theta}{2} \right)} \ket{X_H} + e^{i\phi}\sin{\left(\frac{\theta}{2} \right)} \ket{X_V}
\end{equation}, respectively.
Here $\theta,\phi$ describe the exciton's coherent superposition and are readily identified as the angles describing this  two level system position on its Bloch sphere \cite{winik}. 
Using these projection operators one can express the temporal evolution of the intensity correlation function as:

\begin{equation}\label{e:x-after-xx_new}
    g^{(2)}_{XX_{P_1}-X_{P_2}}\left(\tau > 0 \right) =   \frac{Tr\left(\boldsymbol{\Pi}_{X\left(\theta_2,\phi_2 \right)} e^{\mathcal{L}\tau} \boldsymbol{\Pi}_{X\left(\theta_1,\phi_1 \right)} \right)}{Tr\left (\boldsymbol{\Pi}_{X\left(\theta_2,\phi_2 \right)}\rho_{ss}\right )}
\end{equation}
where $\rho_{ss}$ is the density matrix representing the system's steady state. Note that the long-time evolution always leads to the steady state, therefore $e^{\mathcal{L}\tau} \boldsymbol{\Pi}_{X\left(\theta_1,\phi_1 \right)}\to\rho_{ss} $ and thus the r.h.s. is correctly normalized to $1$ for $\tau=\infty$. 

Freely speaking the equation above describes detection of a biexciton photon with polarization $\varphi(\theta_1, \phi_1)$ which heralds the system in the corresponding coherent superposition of exciton states, then the system evolves for time $\tau$ after which an exciton photon with polarization $\varphi(\theta_2, \phi_2)$, is detected "reading" the exciton state at its annihilation time $\tau$ \cite{bennyprl2011}. 

Since the Lindblad evolution stands for positive times only,  cases where the exciton photon is measured before the biexciton photon ("negative" $\tau$) are described by first projecting on an empty QD state  $\ket{0}$, and second projecting on a biexciton $\ket{XX}$. The intensity correlation function in this case is therefore: 

\begin{equation}\label{e:xx-after-x_new}
    g^{(2)}_{XX_{P_1}-X_{P_2}}\left(\tau < 0 \right) =\frac{
    Tr\, \left(\boldsymbol{\Pi}_{XX} e^{{\mathcal L} |\tau| }\boldsymbol{\Pi}_{0} \right)}{Tr \left(\boldsymbol{\Pi}_{XX}\rho_{ss}\right) }
\end{equation}
Freely speaking here, for a negative $\tau$ correlation event,  the detection of the first photon heralds the state of the QD as empty, and the detection of the second photon "reads" the state of the QD as containing a biexciton.

From the discussion above it follows that there are $4$ QD states, which the measurements of the biexciton-exciton radiative cascade directly probes: $\ket{XX}$, $\ket{X_H}$, $\ket{X_V}$ and $\ket{0}$. 
The system itself, however, may contain very many other additional states such as the dark exciton (DE) \cite{poem2010accessing, schwartz2015deterministic}, multiexcitons \cite{dekel2000cascade} and/or negatively and positively charged excitons and multiexcitons 
 \cite{regelman, benny2012excitation,cygorek2020accurate, cygorek2020atomistic}.
These states should be included in the density matrix which describes the system and the Lindbladian operator should likewise be specified for this density matrix. 
The projection operators one needs to specify, however, are only the above mentioned 4 projection operators. 

Here, for simplicity, we consider only neutral multiexcitons (assuming that the optical excitation leads to QD loading with electron-hole pairs, only). 
In particular we consider the DE, which has equal probability to be photogenerated from an empty QD, as that of the bright exciton (BE). 
The DE radiative recombination rate is very slow \cite{schwartz2015deterministic}, and can be safely 
neglected for generation rates which are typically orders of magnitude larger than the DE decay rate. Higher order metastable dark multiexciton states are also ignored, assuming efficient spin flip processes \cite{benny2014electron, schmidgall2016selection}, which enable multiexcitons' relaxations to their ground energy level. For an even n-multiexciton the ground level is non-degenerate and contain $\frac{n}{2}$ fully occupied electron and hole levels. Therefore the radiative decay rate is uniquely defined. For an odd n-multiexciton there are $ n-1$ electron hole pairs in  $\frac{n-1}{2}$ fully occupied energy levels and the highest energy level contain an unpaired electron and a hole. 
The ground level of odd n-multiexcitons is, therefore, 4-fold degenerate, with 2 dark-like and 2 bright-like states, formed by the unpaired electron-hole spins aligned or anti-aligned.
Since the occupation probability of all 4 states is equal even in cryogenic temperature and since for $n > 1$ both dark- and bright-like excitons are optically active, the radiative rate for these multiexcitons is defined as the average decay rate of both types. 

A schematic diagram of the multiexciton states and the transition rates between these states  is shown in Appendix A Fig. \ref{f:states}.

Generally speaking, if the system is described by $n+3$ states, then the operators will be described by matrices of size $(n+3) \times (n+3)$.
The Hamiltonian of the system is specified by the energies of the various states involved. 
The energies associated with the excitons and multi excitons states are in fact about 4 orders of magnitude larger than the exciton fine structure splitting - $\Delta$. Correspondingly, the relevant time scales associated with the coherent evolution of these levels (optical oscillation times of a few femto- seconds) are far from being resolved in our measurements. Since we are interested in the system evolution on the time scale given by $\Delta$ (about a few hundred picoseconds), it is convenient to remove the fast oscillations associated with the excitonic optical transitions by unitary transformations to the rotating frames with the optical coherent evolution periodicities. Under these transformations the Hamiltonian is indeed independent of the energies of the various multiexciton states and it can be expressed by the projections on the two excitonic eigenstates:

\begin{equation} \label{e:Hamiltonian}
    \mathbf{H} = - \frac{\Delta}{2} \boldsymbol{\Pi}_{X_H} + \frac{\Delta}{2} \boldsymbol{\Pi}_{X_V}
\end{equation}
It is straightforward also to show that the transformations to the rotating frames do not affect the jump operators in the Lindbladian.

The jump operators $\boldsymbol{\Gamma}_{ij}$ must include, however, all the transitions between the various system's states, due to the interactions with the environment (baths). 
We proceed here, for example, following Ref. \cite{dekel2000cascade} by constructing the jump operators assuming a ladder of $n$ neutral multiexcitons, in which the transition rates "up" the ladder are given by the electron-hole pair generation rate $G$ and the transition rates down the ladder are given by each multiexciton-radiative-rate $\frac{1}{\tau_i}$, where $\tau_i$ is the radiative lifetime of the multiexciton state $i$.  
The constructed rate matrix is therefore:
\begin{equation} \label{e:gamma_matrix}
   \boldsymbol{\gamma} =\begin{pmatrix}
        0 & \frac{1}{\tau_H} & \frac{1}{\tau_V} & 0 & 0 & 0 &  &  &  \\
        \frac{G}{4} & 0 & 0 & 0 & \frac{1}{\tau_H} & 0 &  &  &  \\
        \frac{G}{4} & 0 & 0 & 0 & \frac{1}{\tau_V} & 0 &  &  &  \\
        \frac{G}{2} & 0 & 0 & 0 & 0 & 0 &  &  &  \\
        0 & G & G & G & 0 & \frac{1}{\tau_{3}} &  &  &  \\
        0 & 0 & 0 & 0 & G & 0 &  &  &  \\
         &  &  &  &  &  & ... & \frac{1}{\tau_{n-1}} &  \\
         &  &  &  &  &  & G & 0 & \frac{1}{\tau_n} \\
         &  &  &  &  &  &  & G & 0\\
    \end{pmatrix}
\end{equation}

The above mentioned jump operators are therefore constructed from the matrix elements of the rate matrix $\boldsymbol{\gamma}$: 
\begin{equation} \label{e:jump_operators_exp}
    \boldsymbol{\Gamma}_{ij} = \sqrt{\gamma_{ij}} \ket{i} \bra{j}
\end{equation}

We note that the radiative decay times $\tau_i$ can be either directly measured or estimated using simple models \cite{dekel2000cascade,akopian,meirom}. 

The solution of Eq. \ref{e:lindblad-S} for the general case of any Hamiltonian and any rate matrix, is analytically obtained in Appendix \ref{a:AnalyticSol}. 

The solution can be decomposed into two components: the non-coherent one and the one which describes the coherent dynamics of the system. The non-coherent component results from the interactions with the environment, while the coherent component stems from the fine structure splitting between the two exciton's eigenstates.
For the problem constructed by the Hamiltonian from Eq. \ref{e:Hamiltonian} and the rate matrix from Eq. \ref{e:gamma_matrix}, the non-coherent component can be viewed as a solution to the rate equation problem constructed by the diagonal elements of the density matrix, i.e.

\begin{widetext}
\begin{equation}
    \frac{d}{dt} \begin{pmatrix}
        \rho_{00} \vphantom{\frac{1}{\tau_1}} \\
        \rho_{X_H X_H} \vphantom{\frac{1}{\tau_1}} \\
        \rho_{X_V X_V} \vphantom{\frac{1}{\tau_1}} \\
        \rho_{X_{DE} X_{DE}} \vphantom{\frac{1}{\tau_1}} \\
        \rho_{X\!X\ X\!X} \vphantom{\frac{1}{\tau_1}} \\
        \rho_{33} \vphantom{\frac{1}{\tau_1}} \\
        ... \vphantom{\frac{1}{\tau_1}}\\
        \rho_{ii} \vphantom{\frac{1}{\tau_1}}\\
        ... \vphantom{\frac{1}{\tau_1}}\\
        \rho_{nn} \vphantom{\frac{1}{\tau_1}}
    \end{pmatrix}
    = \begin{pmatrix}
        -G & \frac{1}{\tau_H} & \frac{1}{\tau_V} & 0 & 0 & 0 & & & \\
        \frac{G}{4} & -G -\frac{1}{\tau_H} & 0 & 0 & \frac{1}{\tau_H} & 0 & & & \\
        \frac{G}{4} & 0 & -G -\frac{1}{\tau_V} & 0 & \frac{1}{\tau_V} & 0 &  & & \\
        \frac{G}{2} & 0 & 0 & -G & 0 & 0 & & &  \\
        0 & G & G & G & -G -\frac{1}{\tau_H} -\frac{1}{\tau_V} & \frac{1}{\tau_3} & & & \\
        0 & 0 & 0 & 0 & G & -G-\frac{1}{\tau_3} & & & \\
         &  &  &  &  &  & ... & \frac{1}{\tau_i} & & \\
         &  &  &  &  &  & G & -G-\frac{1}{\tau_i} & & \\
         &  &  &  &  &  &  &  & ... & \frac{1}{\tau_n}\\
         &  &  &  &  &  &  &  & G & -\frac{1}{\tau_n}\\
    \end{pmatrix}
    \cdot 
    \begin{pmatrix}
        \rho_{00} \vphantom{\frac{1}{\tau_1}} \\
        \rho_{X_H X_H} \vphantom{\frac{1}{\tau_1}} \\
        \rho_{X_V X_V} \vphantom{\frac{1}{\tau_1}} \\
        \rho_{X_{DE} X_{DE}} \vphantom{\frac{1}{\tau_1}} \\
        \rho_{X\!X\ X\!X} \vphantom{\frac{1}{\tau_1}} \\
        \rho_{33} \vphantom{\frac{1}{\tau_1}} \\
        ... \vphantom{\frac{1}{\tau_1}}\\
        \rho_{ii} \vphantom{\frac{1}{\tau_1}}\\
        ... \vphantom{\frac{1}{\tau_1}}\\
        \rho_{nn} \vphantom{\frac{1}{\tau_1}}
    \end{pmatrix}
\end{equation}
\end{widetext}

The solution to this system is a sum of eigenvectors of the matrix, each evolving as an exponential term with equivalent eigenvalue, as in \cite{regelman}. Notably, these eigenvalues are real, yielding a transient solution that lacks oscillatory behavior.

Oscillations in the full solution, however, stem from the coherent part, which introduces imaginary contributions to the eigenvalues associated with the excitonic states.

Explicitly, the Hamiltonian part influences the two exciton's off-diagonal terms in the density matrix, such that they will evolve as:
\begin{equation}
\begin{aligned}
    \boldsymbol{\rho}_{X_H X_V}(t)
    &= \boldsymbol{\rho}_{X_H X_V}(0) e ^ {\left( \frac{i\Delta} {\hbar} - \frac{1}{\tau_1} - G \right) t} \\
    \boldsymbol{\rho}_{X_V X_H}(t)
    &= \boldsymbol{\rho}_{X_V X_H}(0) e ^ {\left( -\frac{i\Delta} {\hbar} - \frac{1}{\tau_1} - G \right) t}
\end{aligned}
\end{equation}
This aspect is detailed in Appendix \ref{a:AnalyticSol}.

The ability to separate between the coherent and incoherent components of the solution for the dynamics of the system was used previously in analyzing experimental studies of the biexciton-exciton radiative cascade \cite{akopian}. Akopian et al, subtracted the pure incoherent measurement (cross rectilinearly polarized biexciton-exciton photon pairs) from the experimental data which included also coherent dynamics. 
This yielded a very good approximation for the coherent dynamics of the cascading photons, which in turn permitted the first measurement of the degree of entanglement between the two photons.  

\section{Results}

With the approximations discussed above, the model is fully defined by the set of parameters $\{ \Delta, \tau_i, G \}$. The parameters $\{ \Delta, \tau_i \}$ can be independently determined experimentally, the first by the spectral measurement of the exciton FSS and the rest using time resolved PL measurements of identified multiexciton spectral lines. Decay times of high order multiexciton lines, which are not readily identified spectrally, can be estimated by using models \cite{dekel2000cascade,akopian,meirom}.
The electron-hole generation rate $G$ is in general proportional to the excitation intensity, and thus can be quite accurately determined by fitting the model to two or more sets of polarization sensitive correlation measurements under various excitation intensities (not shown in this work).  In the following, we left $G$ as a free fitting parameter.

The use of two LCVRs for polarization projection is extremely convenient from the experimental point of view. Its calibration, however is not straight forward and it may introduce systematic deviations in the angles 
$\theta$ and $\phi$ as defined in Eqs. 4 and 5. Photonic nanostructures such as micropillars, nanowires and/or circular Bragg reflectors may also contribute to these systematic deviations.
We define the systematic deviations as $\Delta{\theta}=\Delta{\theta_1}=\Delta{\theta_2}$ and $\Delta{\phi}=\Delta{\phi_1}=\Delta{\phi_2}$, and used them as parameters in the actual fitting procedure
 $\boldsymbol{\Pi}_{X\left(\theta_{i}+\Delta\theta,\phi_{i}+\Delta{\phi} \right)}$.
Using these 3 fitting parameters we quite successfully fitted all 36 polarization sensitive time resolved biexciton-exciton correlation measurements.

\begin{figure}
  \centering
  \subfloat[]{\includegraphics[width=0.9\columnwidth]{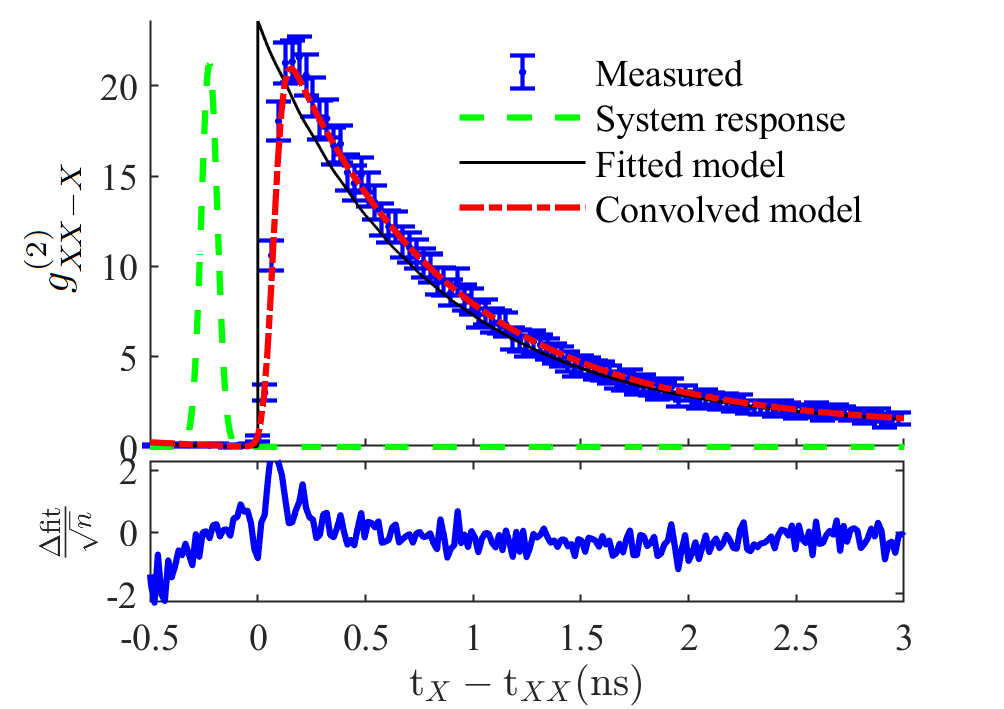}}
  \hfill
  \subfloat[]{\includegraphics[width=0.9\columnwidth]{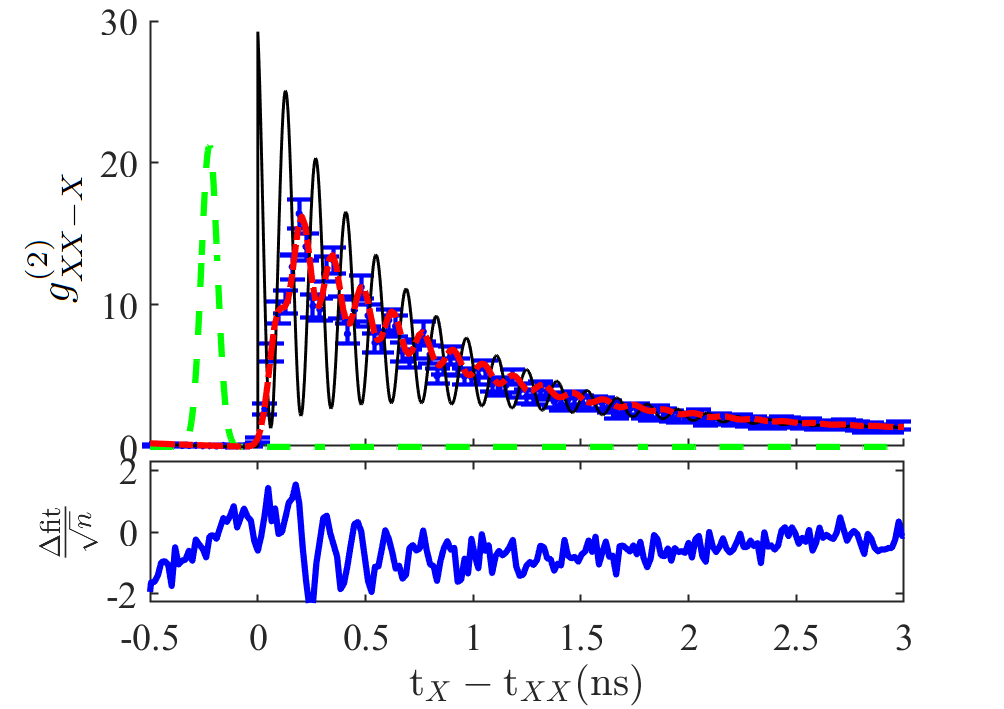}\label{f:RR}}
  \hfill
  \subfloat[]{\includegraphics[width=0.9\columnwidth]{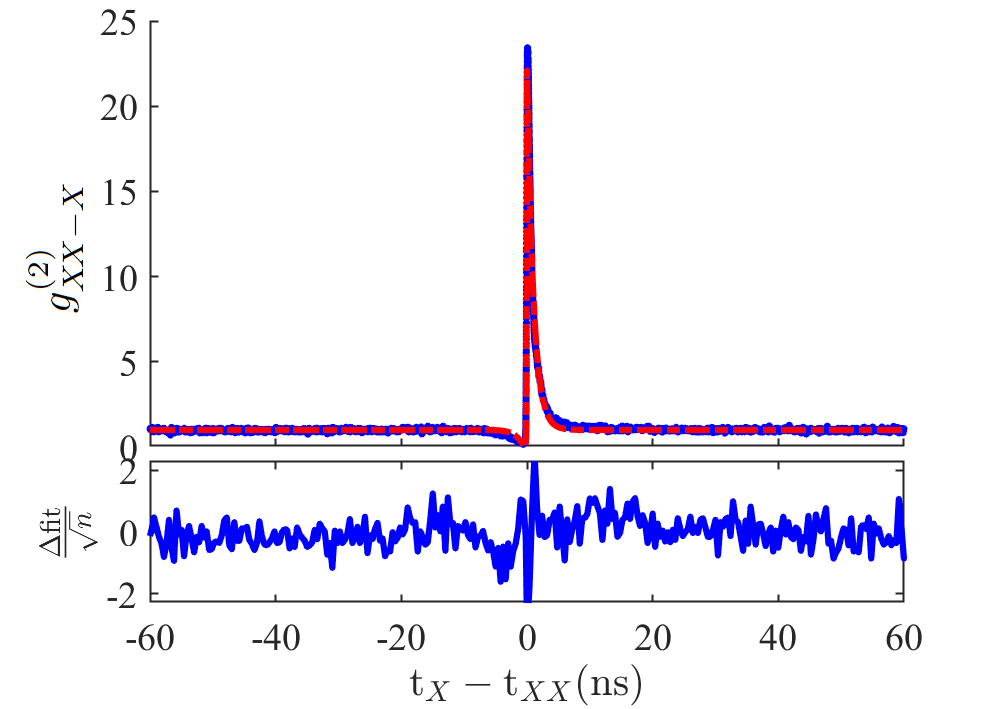}\label{f:HH_long}}
  \caption{\justifying Polarization sensitive time resolved intensity correlation measurements  (blue data points and error bars), the best fitted model calculation (black solid line) and the model convolved with the detector response function (red dashed line). The green line shows the detector's response function used for convolution. (a) [(b)] shows the case where both photons are co-linearly [-circularly] polarized H [R]. The sub-panel below the figure shows the difference between the measured data and the fitted convolved model normalized by the experimental uncertainty. (c) shows the measured and best fitted intensity correlation function (Eq. \ref{e:intensity-correlation}) for the same case as (a) but for long times, demonstrating the return to steady state  
($ g^{(2)}\left(\tau\rightarrow\pm\infty\right)=1$).
The measured 2-photons coincidence rate was about  $10$ KHz. }\label{f:fit}
\end{figure}

In Fig.~\ref{f:fit} we show two typical time resolved measurements from the whole set of 36 measurements. In one measurement the two photons are co-rectilinearly H polarized (Fig. \ref{f:fit}(a)) and in the other (Fig. \ref{f:fit}(b)) the two photons are co-circularly  R polarized. The blue dots stand for the measured data, the error bars represent one standard deviation, and the black solid line stands for the best fitted Eqs.~\ref{e:x-after-xx_new} and \ref{e:xx-after-x_new} to the measured data. For the fitting we used the measured excitonic FSS of $\Delta=29 \mu eV$ which resulted in a precession period of $\frac{\hbar}{\Delta} = 140 \pm 10$ ps.
We found that the measured lifetimes of the QD confined exciton's  eigenstates \cite{dekelssc2001} are not equal with $\tau_H = 1180 \pm 10$ ps and $\tau_V = 990 \pm 10$ ps. The difference is probably due to different coupling strengths to the nanowire optical mode.

Radiative lifetimes on the order of $1$ ns are quite typical for these types of quantum dots. 
Though calculating these lifetimes reliably, require exact knowledge of the QD and nanowire dimensions, composition profiles and the exact position of the QD within the nanowire, we suggest here a simple microscopic   model that seems to reproduce the measured lifetime.
For a spherical QD one gets \cite{dekelssc2001}:
\begin{equation}\label{eq:tau-r}
\frac{1}{\tau_r}=\frac{4e^2k_0^2f}{n_mm_0c}
\end{equation}
Where $k_0=(n_mE_{ex})/(\hbar c)$   is the minimal photon $k-vector$, $E_{ex}=1.283 eV$ is the exciton energy, $n_m=3.12$ is the nanowire material index of refraction, $e$ and $m_0$ are the electronic charge and mass, $\hbar$ is the reduced Planck constant, c  the speed of light and $f \approx 1$ is the unitless oscillator strength given by the overlap between the electron and hole envelope wavefunctions \cite{citrin1993radiative, andreani1991radiative}.
Using Eq. \ref{eq:tau-r} we get $\tau_r= 2.25$ ns. 
The measured lifetime of about a factor of 2 shorter can be explained by the reduction of the mode volume enforced by the nanowire of subwavelength diameter ($d_w=200$ nm). The lifetime shortening should be given by the square of the ratio between the wire diameter and the photon wavelength in matter $\lambda_m=\frac{2\pi\hbar c}{E_{ex} n_m}\approx 310$ nm: 
\begin{equation}\label{eq:tau-x}
\tau_x=\tau_r(\frac{d_w}{\lambda_m})^2=0.95\  \text{ns}
\end{equation}

For fitting the experimental measurements, we used the measured lifetimes for the biexciton and exciton optical transitions as depicted in Fig. \ref{f:cascade}. For higher order multiexcitons, which become increasingly important as one raises the excitation intensity (and therefore G), we follow Ref. \cite{regelman}  and define the radiative rate of the n-order multiexciton as given by the number of available radiative recombination channels to the multiexciton of order n-1. Each such channel involves annihilation of an electron-hole pair with opposite spin projections on the direction of the light optical direction. 
Thus, for example, the radiative lifetime of the biexciton (multiexciton of order n=2) is approximately half the radiative lifetime of the exciton (multiexciton of order n=1), since there are 2 allowed radiative channels for its decay. Similarly, for higher n-multiexcitons of even order, the number of allowed radiative recombination channels is given by 2 from each fully occupied level. Since the number of occupied levels is $\frac{n}{2}$, the number of allowed optical transitions is exactly n. Obviously, this is also the case for bright-like odd n-multiexcitons, while dark-like odd n-multiexcitons, have  only $(n-1)$ transitions. Therefore, the average number of allowed recombination channels for odd n multiexcitons is $n-\frac{1}{2}$.

In summary we used:
\begin{equation}\label{eq:tau-i}
\tau_i=
    \begin{cases}
        \tau_X/(i- \frac{1}{2}) & i \text{ is Odd} \\
        \tau_X/i & i \text{ is Even}
    \end{cases} \ \ \ \ , \ \ \ \  i\geq3
\end{equation}
with $\frac{1}{\tau_X} = \frac{1}{2}\left(\frac{1}{\tau_H} + \frac{1}{\tau_V} \right)$ is the mean exciton's decay rate.

The inclusion of higher order multiexcitons is required when one considers the system under strong excitation. Obviously, the stronger the excitation is the higher is the confined level that carriers occupy (due to the Pauli exclusion principle). The highest order multiexciton that one chooses to consider, depends on the probability to find such a multiexciton in the QD at steady state. This probability can be easily calculated by our model.  In practice, we increased $n$ until there was no longer increase in the quality of the fits that we produced, and then checked the actual occupancy of the $n^{th}$-multiexciton level, for consistency.

The red dashed line in Fig.~\ref{f:fit} represents the best fitted model, convoluted with the temporal response function of the experimental system as represented by the green dashed line.  
In order to simplify the convolution procedure and make it analytic, we approximated the response function by a Gaussian function with full width at half maximum of $42$ ps.
For the particular fitting in Fig.~\ref{f:fit}, we used $\frac{1}{G} = 8.0 \pm 0.5$ ns, $\Delta\theta = (0.10\pm 0.02)\pi$ and $\Delta\phi = (0.02\pm 0.02)\pi$. 
The low panels in Fig.~\ref{f:fit} show the time resolved differences between the measured data and the best fitted model, normalized by the experimental uncertainty of the measurements. 
In Fig. \ref{f:fit}(c) we present the data for an extended time scale, long after the system reaches steady state.

Fig.~\ref{f:fit}, demonstrates that our Lindblad model fits the data quite well. 
For short times and low generation rates, the results align closely with those of the Hamiltonian model used for pulse excitation \cite{winik}. 
For long times, the coherent dynamics loses significance, and the measured results are similar to those described by the non-coherent rate equation model \cite{regelman}.

We note that the best fitted value of G, obtained from the time resolved measurements is also in agreement with the measured steady state ratio between the intensities of the biexciton and exciton spectral lines (about 0.65) as shown in Fig. \ref{f:experiment}(b). 
Under these conditions, the steady state occupation probabilities were 0.597 for the empty QD, 0.298 for the DE, 0.039 for each of the BE states ($X_H$ and $X_V$), 0.025 for the XX, and $3 \cdot 10^{-4}$, $1 \cdot 10^{-5}$, and $1 \cdot 10^{-7}$ for the $n=3$, $n=4$, and $n=5$ multiexciton states, respectively. Higher order multiexciton occupations were orders of magnitude smaller and their inclusion did not improve the quality of the fits to the experimental data.

\begin{figure*}[]
    \includegraphics[width=1\textwidth]{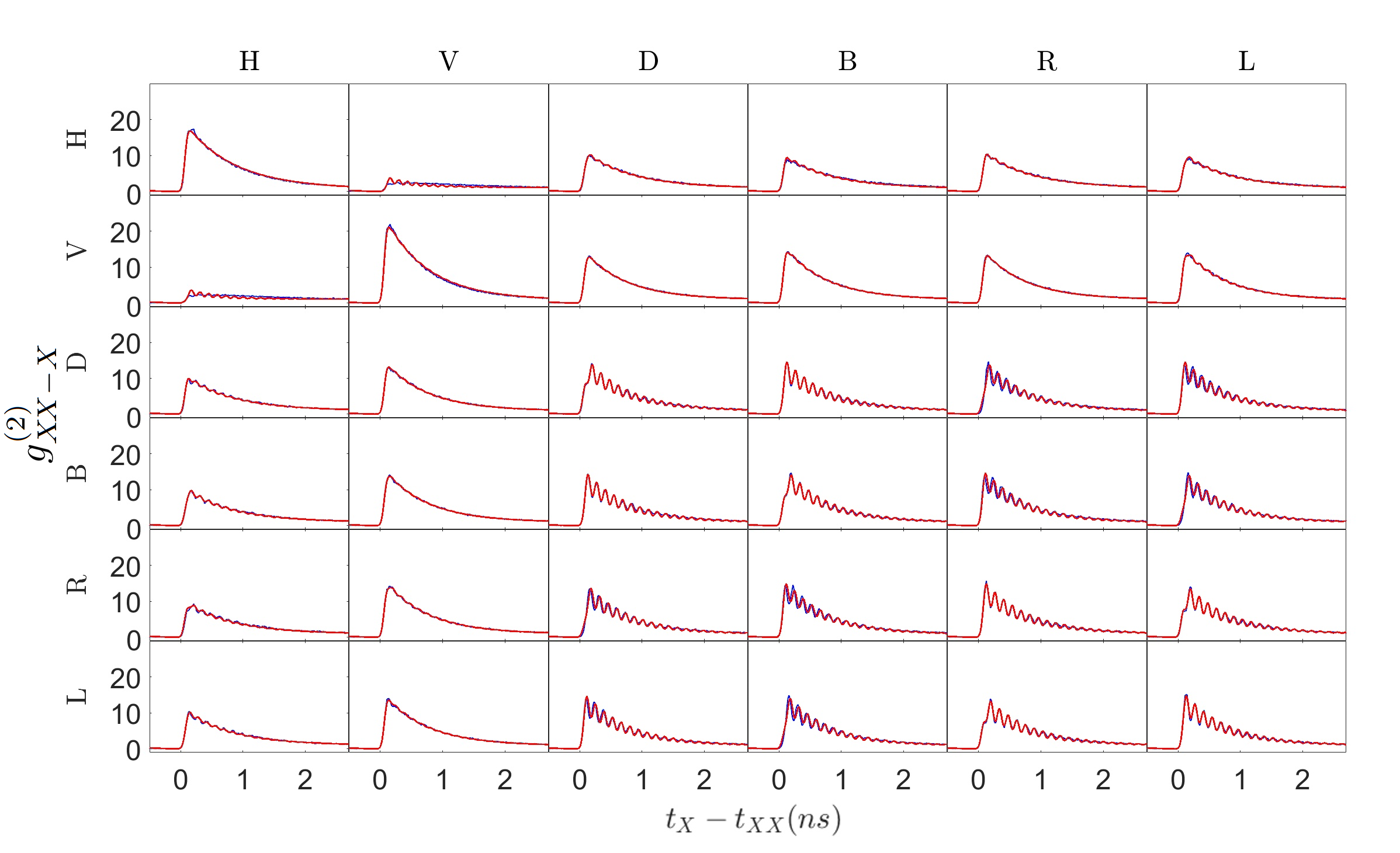}
    \caption{\justifying 36 different polarization sensitive time resolved 2-photon correlation measurements (solid blue lines) and best fitted convolved model (solid red line) . The entry $P_1P_2$ in the table gives the probability of detecting a $P_2$-polarized exciton photon at time $\tau>0$, conditioned on the detection of a $P_1$ polarized biexciton photon at time 0.} \label{f:36}
\end{figure*}

Fig.~\ref{f:36} shows 36 polarization-sensitive correlation measurements. The measurements are in blue and the fitted convolved models are in red. 

\section{Summary}

We studied experimentally and theoretically the polarization-sensitive intensity cross-correlation functions of the QD confined biexciton-exciton cascade for a system driven by a non-resonant CW excitation. 
The CW excitation is modeled by an electron-hole pair bath that feeds the QD. 
The system temporal evolution is described by a set of Lindbladian equations.
The time resolved intensity cross-correlation functions for 36 different polarization sensitive measurements have been fitted quite successfully by our model using a minimal set of fitting parameters. 
The theoretical framework that we outlined here can be readily extended to include additional coherent multiexciton states 
while preserving the simplicity and efficiency of the solution method.

\section*{Acknowledgments}
This research was supported by the Israeli Science Foundation (ISF - grant No. 1933/23) the European Research Council (ERC-Grant No. 695188) and the German Israeli Research Cooperation (DIP - grant No. DFG-FI947-6-1).

We thank Raz Firenko, Klaus Molmer and Pawel Hawrylak for useful discussions.

\clearpage
\appendix

\section{Analytic solution to the Lindblad equations} \label{a:AnalyticSol}
This Appendix provides a full analytic solution for a general case of the Lindblad equations shown in the main paper.

\subsection{General form of the differential equations}
As shown in Eq. \ref{e:lindblad-S}, the Lindblad equation consists of two terms. The first, ${\cal H}(\boldsymbol{\rho})$, describes the Hamiltonian evolution and the second, ${\cal D}(\boldsymbol{\rho})$, describes the interactions with the environment. In this Appendix, we present the general form of the Lindblad equation for the biexciton-exciton cascade, and later on we will present its full analytic solution for a specific case.

The Hamiltonian can be written in it's diagonal form using the energy $E_i$ of each state, such that a general form of it is:
\begin{equation}
    \mathbf{H} = \sum_{i} E_i \ket{i}\bra{i}
\end{equation}
and therefore the Hamiltonian term in the Lindblad equation takes the form:
\begin{equation}
\begin{aligned}
    {\cal H}(\boldsymbol{\rho})
    &= -\frac{i}{\hbar}\left[ \mathbf{H},\boldsymbol{\rho} \right] \\
    &= \frac{i}{\hbar} \left( \boldsymbol{\rho}\mathbf{H} - \mathbf{H}\boldsymbol{\rho} \right) \\
    &= \frac{i}{\hbar} \sum_{i} E_i \left( \boldsymbol{\rho} \ket{i}\bra{i} - \ket{i}\bra{i}\boldsymbol{\rho} \right)
\end{aligned}
\end{equation}
which leads to a matrix element of:
\begin{equation}
\begin{aligned}
    \bra{a} {\cal H}(\boldsymbol{\rho}) \ket{b}
    &= \frac{i}{\hbar} \sum_{i} E_i \left( \bra{a} \boldsymbol{\rho} \ket{i}\bra{i}\ket{b} - \bra{a}\ket{i}\bra{i}\boldsymbol{\rho} \ket{b} \right) \\
    &= \frac{i}{\hbar} \sum_{i} E_i \left( \boldsymbol{\rho}_{ai} \delta_{ib} - \delta_{ai} \boldsymbol{\rho}_{ib} \right) \\
    &= \frac{i}{\hbar} \left( E_b - E_a \right)\boldsymbol{\rho}_{ab}
\end{aligned}
\end{equation}

The second part of the general Lindblad equation is a sum of terms:
\begin{equation}
    {\cal D}(\boldsymbol{\rho}) 
    = \sum_{ij} {\cal D}_{ij}(\boldsymbol{\rho})
\end{equation}
where each term has the form of Eq. \ref{e:superHsuperD}, and the general form of the jump operators is defined by a rate matrix - $\boldsymbol{\gamma}$, such that:
\begin{equation}
    \boldsymbol{\Gamma}_{ij} = \sqrt{\gamma_{ij}} \ket{i}\bra{j}
\end{equation}
as in Eq. \ref{e:jump_operators_exp}. Using these general definitions, one can write the second part of the Lindblad equation as:
\begin{equation}
\begin{aligned}
    {\cal D}(\boldsymbol{\rho}) 
    &= \frac{1}{2} \sum_{ij} \left( [\boldsymbol{\Gamma}_{ij} \boldsymbol{\rho}, \boldsymbol{\Gamma}_{ij} ^\dagger] + [\boldsymbol{\Gamma}_{ij} ,\boldsymbol{\rho} \boldsymbol{\Gamma}_{ij} ^\dagger] \right) \\
    &= \sum_{ij} \left( \boldsymbol{\Gamma}_{ij} \boldsymbol{\rho} \boldsymbol{\Gamma}_{ij} ^\dagger
    - \frac{1}{2} \boldsymbol{\Gamma^\dagger}_{ij} \boldsymbol{\Gamma}_{ij}\boldsymbol{\rho}
    - \frac{1}{2} \boldsymbol{\rho} \boldsymbol{\Gamma^\dagger}_{ij} \boldsymbol{\Gamma}_{ij} \right) \\
    &= \sum_{ij} \gamma_{ij} \left( \ket{i}\bra{j} \boldsymbol{\rho} \ket{j}\bra{i}
    - \frac{1}{2} \ket{j}\bra{j} \boldsymbol{\rho}
    - \frac{1}{2} \boldsymbol{\rho} \ket{j}\bra{j} \right)
\end{aligned}
\end{equation}
which leads to a matrix element of:
\begin{equation}
\begin{aligned}
    \bra{a} {\cal D}(\boldsymbol{\rho}) \ket{b}
    &= \sum_{ij} \gamma_{ij} \biggl( \bra{a}\ket{i}\bra{j} \boldsymbol{\rho} \ket{j}\bra{i} \ket{b}  \\
    &- \ \ \  \frac{1}{2} \bra{a}\ket{j}\bra{j} \boldsymbol{\rho} \ket{b}
    - \frac{1}{2} \bra{a}\boldsymbol{\rho} \ket{j}\bra{j} \ket{b} \biggr) \\
    &= \sum_{ij} \gamma_{ij} \left( \delta_{ai} \boldsymbol{\rho}_{jj} \delta_{ib}
    - \frac{1}{2} \delta_{aj} \boldsymbol{\rho}_{jb}
    - \frac{1}{2} \boldsymbol{\rho}_{aj} \delta_{jb} \right) \\
    &=  \sum_{i} \left( \gamma_{ai} \delta_{ab} \boldsymbol{\rho}_{ii}
    - \frac{1}{2} \left( \gamma_{ia} + \gamma_{ib} \right) \boldsymbol{\rho}_{ab} \right)
\end{aligned}
\end{equation}
Finally by combining the elements one gets the full general differential equation for density matrix element $\boldsymbol{\rho}_{ab}$:
\begin{equation}
\begin{aligned}
    \frac{d}{dt}\boldsymbol{\rho}_{ab}
    &= \frac{i}{\hbar}\left(E_b - E_a \right) \boldsymbol{\rho}_{ab} \\
    &+ \sum_{i} \left( \gamma_{ai} \delta_{ab} \boldsymbol{\rho}_{ii}
    - \frac{1}{2} \left( \gamma_{ia} + \gamma_{ib} \right) \boldsymbol{\rho}_{ab} \right)
\end{aligned}
\end{equation}

\subsection{General solution}
The equations have different form for diagonal and non-diagonal terms of the density matrix. For diagonal terms the equations for $\rho_{ab}$ decouple and take the form:
\begin{equation}\label{a:general_non_coherent}
    \frac{d}{dt}\boldsymbol{\rho}_{aa}
    = \sum_{i} \left( \gamma_{ai} \boldsymbol{\rho}_{ii}
    - \gamma_{ia} \boldsymbol{\rho}_{aa} \right)
\end{equation}
which has exactly the same form of the well known \cite{regelman} rate equation constructed by the diagonal terms of the density matrix, and the rate matrix - $\boldsymbol{\gamma}$.

For non-diagonal terms the equation takes the form of:
\begin{equation}
    \frac{d}{dt}\boldsymbol{\rho}_{ab}
    = \left( \frac{i}{\hbar}\left(E_b - E_a \right) 
    - \frac{1}{2} \sum_{i} \left( \gamma_{ia} + \gamma_{ib} \right) \right) \boldsymbol{\rho}_{ab}
\end{equation}
which is a simple differential equation, and it's closed solution is:
\begin{equation} \label{a:general_coherent}
    \boldsymbol{\rho}_{ab}(t) = \boldsymbol{\rho}_{ab}(0) e^{\left( \frac{i}{\hbar}\left(E_b - E_a \right) 
    - \frac{1}{2} \sum_{i} \left( \gamma_{ia} + \gamma_{ib} \right) \right) t }
\end{equation}

\subsection{Case-Specific Derivation}
Using the general solution derived in the previous section, specifically Eqs.~\ref{a:general_non_coherent} and \ref{a:general_coherent}, one can obtain the solution for the specific scenario considered in this paper as follows:

By substituting the $\boldsymbol{\gamma}$ matrix defined in Eq.~\ref{e:gamma_matrix} into Eq.~\ref{a:general_non_coherent}, the diagonal terms of the density matrix, corresponding to the non-coherent part of the solution are determined. These terms reduce to the well-known rate equations for multiexciton systems \cite{regelman}:
\begin{equation}
    \frac{d}{dt}\boldsymbol{\rho}_{ii}
    = G \boldsymbol{\rho}_{(i-1)(i-1)} + \frac{1}{\tau_{(i+1)}}\boldsymbol{\rho}_{(i+1)(i+1)}
    - G \boldsymbol{\rho}_{ii}
    - \frac{1}{\tau_{i}} \boldsymbol{\rho}_{ii}
\end{equation}

The non-diagonal terms, corresponding to the coherent part, described by Eq.~\ref{a:general_coherent}, are expressed as simple exponential functions. In the specific scenario discussed in this paper, coherence is present only between the excitonic states. Consequently, the only relevant non-diagonal terms are $\boldsymbol{\rho}_{X_H X_V}$ and $\boldsymbol{\rho}_{X_V X_H}$, which, using the $\boldsymbol{\gamma}$ matrix and the FSS - $\Delta$, evolve as:
\begin{equation}
\begin{aligned}
    \boldsymbol{\rho}_{X_H X_V}(t)
    &= \boldsymbol{\rho}_{X_H X_V}(0) e ^ {\left( \frac{i\Delta} {\hbar} - \frac{1}{\tau_1} - G \right) t} \\
    \boldsymbol{\rho}_{X_V X_H}(t)
    &= \boldsymbol{\rho}_{X_V X_H}(0) e ^ {\left( -\frac{i\Delta} {\hbar} - \frac{1}{\tau_1} - G \right) t}
\end{aligned}
\end{equation}

 This completes the derivation of the model used in the paper.

\section{ Schematic description of the multiexciton states}
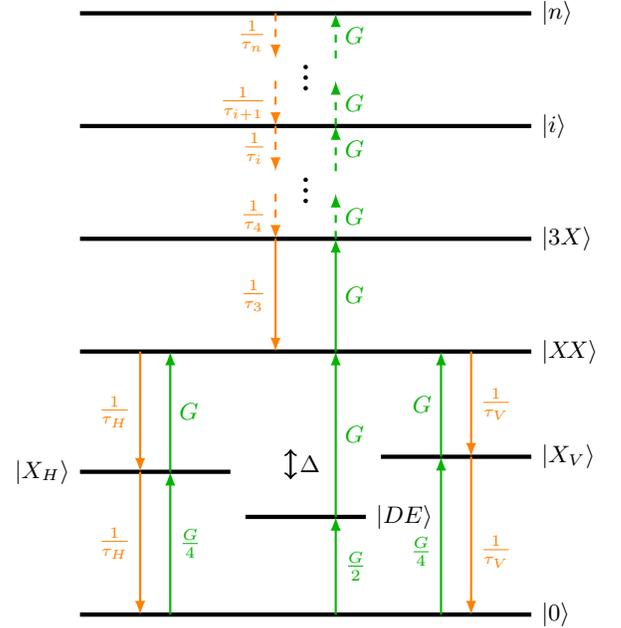
\begin{figure}[ht]
    \centering
    \begin{tikzpicture}[
        every node/.style={font=\small},
        level/.style={ultra thick},  
        trans/.style={->, >=latex, thick},
        orangetrans/.style={->, >=latex, thick, orange},
        orangedashedtrans/.style={->, >=latex, thick, orange, dashed},
        greentrans/.style={->, >=latex, thick, green!70!black},
        greendashedtrans/.style={->, >=latex, thick, green!70!black, dashed}
    ]
        \draw[level] (-3, -2) -- (3, -2) node[right] {\(|0\rangle\)};
        \draw[level] (-0.8, -0.7) -- (0.8, -0.7) node[right] {\(|DE\rangle\)};
        \draw[level] (-1, -0.1) -- (-3, -0.1) node[left] {\(|X_H\rangle\)};
        \draw[level] (1, 0.1) -- (3, 0.1) node[right] {\(|X_V\rangle\)};
        \draw[level] (-3, 1.5) -- (3, 1.5) node[right] {\(|X\!X\rangle\)};
        \draw[level] (-3, 3) -- (3, 3) node[right] {\(|3X\rangle\)};
        \draw[level] (-3, 4.5) -- (3, 4.5) node[right] {\(|i\rangle\)};
        \draw[level] (-3, 6) -- (3, 6) node[right] {\(|n\rangle\)};

        \draw[greentrans] (0.4, -2) -- (0.4, -0.7) node[midway, right, green!70!black] {\(\frac{G}{2}\)}; 
        \draw[greentrans] (1.8, -2) -- (1.8, 0.1) node[pos=0.4, left, green!70!black] {\(\frac{G}{4}\)}; 
        \draw[greentrans] (-1.8, -2) -- (-1.8, -0.1) node[midway, right, green!70!black] {\(\frac{G}{4}\)}; 
        \draw[greentrans] (1.8, 0.1) -- (1.8, 1.5) node[midway, left, green!70!black] {\(G\)}; 
        \draw[greentrans] (-1.8, -0.1) -- (-1.8, 1.5) node[midway, right, green!70!black] {\(G\)}; 
        \draw[greentrans] (0.4, -0.7) -- (0.4, 1.5) node[midway, right, green!70!black] {\(G\)}; 
        \draw[greentrans] (0.4, 1.5) -- (0.4, 3) node[midway, right, green!70!black] {\(G\)}; 
        \draw[greendashedtrans] (0.4, 3) -- (0.4, 3.6) node[midway, right, green!70!black] {\(G\)}; 
        \draw[greendashedtrans] (0.4, 3.9) -- (0.4, 4.5) node[midway, right, green!70!black] {\(G\)}; 
        \draw[greendashedtrans] (0.4, 4.5) -- (0.4, 5.1) node[midway, right, green!70!black] {\(G\)}; 
        \draw[greendashedtrans] (0.4, 5.4) -- (0.4, 6) node[midway, right, green!70!black] {\(G\)}; 

        \draw[orangedashedtrans] (-0.4, 6) -- (-0.4, 5.4) node[midway, left, orange] {\(\frac{1}{\tau_{n}}\)}; 
        \draw[orangedashedtrans] (-0.4, 5.1) -- (-0.4, 4.5) node[midway, left, orange] {\(\frac{1}{\tau_{i+1}}\)}; 
        \draw[orangedashedtrans] (-0.4, 4.5) -- (-0.4, 3.9) node[midway, left, orange] {\(\frac{1}{\tau_i}\)}; 
        \draw[orangedashedtrans] (-0.4, 3.6) -- (-0.4, 3) node[midway, left, orange] {\(\frac{1}{\tau_4}\)}; 
        \draw[orangetrans] (-0.4, 3) -- (-0.4, 1.5) node[midway, left, orange] {\(\frac{1}{\tau_3}\)}; 
        \draw[orangetrans] (-2.2, 1.5) -- (-2.2, -0.1) node[midway, left, orange] {\(\frac{1}{\tau_H}\)}; 
        \draw[orangetrans] (-2.2, -0.1) -- (-2.2, -2) node[midway, left, orange] {\(\frac{1}{\tau_H}\)}; 
        \draw[orangetrans] (2.2, 1.5) -- (2.2, 0.1) node[midway, right, orange] {\(\frac{1}{\tau_V}\)}; 
        \draw[orangetrans] (2.2, 0.1) -- (2.2, -2) node[pos=0.6, right, orange] {\(\frac{1}{\tau_V}\)}; 

        \draw[<->, thick] (-0.2, -0.2) -- (-0.2, 0.2) node[midway, right] {\(\Delta\)};
        \node at (0, 3.75) {\huge \(\boldsymbol{\vdots}\)};
        \node at (0, 5.25) {\huge \(\boldsymbol{\vdots}\)};

    \end{tikzpicture}
    \caption{\justifying{
    States Diagram. Black horizontal lines represent the various levels. Arrows between levels represent transitions between them. Orange (green) arrows represent optical electron-hole pair recombination (generation).
    }}
    \label{f:states}
\end{figure}

In Fig. \ref{f:states} we present a diagram which schematically describes the multiexciton state ladder and the transition rates between these states. These states and rates were considered in the specific example discussed in this paper.

\bibliography{references}

\end{document}